\documentclass[aps,superscriptaddress,showkeys,showpacs,twocolumn]{revtex4}%

\usepackage{amssymb}
\usepackage{amsmath}
\usepackage{graphicx}
\usepackage[normalem]{ulem}
\usepackage[dvips]{color}
\usepackage{CJK}                     
\usepackage[normalem]{ulem} 
\usepackage[dvips]{color} 
\renewcommand\sout{\bgroup \color{red} \ULdepth=-.5ex \ULset}

\def\rpi {$\pi^-/\pi^+$~}
\def\es0{$E_{sym}(\rho_0)$~}
\begin{document}
\begin{CJK*}{GBK}{song}
\title{Impact parameter dependence of \rpi ratio in probing the nuclear symmetry energy using heavy-ion collisions}
\author{Gao-Feng Wei}\email[Corresponding author: ]{wei.gaofeng@foxmail.com}
\affiliation{Shaanxi Key Laboratory of Surface Engineering and Remanufacturing, School of Mechanical and Material Engineering, Xi'an University, Xi'an, 710065, China}
\affiliation{Department of Physics and Astronomy, Texas A$\&$M University-Commerce, Commerce, TX 75429-3011, USA}
\author{Guo-Qiang He}
\affiliation{School of Electronic Engineering, Xi'an Shiyou University, Xi'an, 710065, China}
\author{Xin-Wei Cao}
\affiliation{Shaanxi Key Laboratory of Surface Engineering and Remanufacturing, School of Mechanical and Material Engineering, Xi'an University, Xi'an, 710065, China}
\author{Yi-Xin Lu}
\affiliation{Shaanxi Key Laboratory of Surface Engineering and Remanufacturing, School of Mechanical and Material Engineering, Xi'an University, Xi'an, 710065, China}

\begin{abstract}
The impact parameter dependence of \rpi ratio is examined in heavy-ion collisions at 400MeV/nucleon within
a transport model. It is shown that the sensitivity of \rpi ratio on symmetry energy shows a transition
from central to peripheral collisions, i.e., the stiffer symmetry energy leads to a larger \rpi
ratio in peripheral collisions while the softer symmetry energy always leads this ratio to be larger in
central collisions. After checking the kinematic energy distribution of \rpi ratio, we found this transition
of sensitivity of \rpi ratio to symmetry energy is mainly from less energetic pions, i.e., the softer symmetry
energy gets the less energetic pions to form a smaller \rpi ratio in peripheral collisions while these pions
generate a larger \rpi ratio in central collisions. Undoubtedly, the softer symmetry energy can also lead more
energetic pions to form a larger \rpi ratio in peripheral collisions. Nevertheless, considering that most of
pions are insufficient energetic at this beam energy, we therefore suggest the \rpi ratio as a probe of the
high-density symmetry energy effective only in central at most to midcentral collisions, thereby avoiding the
possible information of low-density symmetry energy carried in \rpi ratio from peripheral collisions.

\end{abstract}

\maketitle

The determination of density-dependent nuclear symmetry energy is one of the hot topic
in isospin physics due to its importance in understanding the structure of radiative
nuclei in nuclear physics \cite{Oya98,Bro00,Hor01,Fur02} and the evolution of massive
stars and properties of neutron stars in nuclear astrophysics
\cite{Bet90,Lat01,Eng94,Pra88,Che03a}. Presently, although many useful experimental
observables \cite{Bao04,Mul95,Bar02,Che03b,Shi03,Sca99,Bao01,Bao02,Gai04,Dan98} have
been proposed to determine the nuclear symmetry energy, the knowledge regarding the nuclear
symmetry energy is still far lack except for the relative determination of nuclear symmetry
energy at saturation density $\rho_{0}$ from empirical liquid-drop mass formula \cite{Mye66,Che03b}.
For example, by comparing the \rpi ratio with the FOPI experimental data \cite{FOPI}, the
Boltzmann-Uehling-Uhlenbeck (BUU) \cite{Xiao09} and Boltzmann-Langevin (BL) \cite{Xie13}
communities favor a super-soft symmetry energy, but the quantum molecular dynamics (QMD)
\cite{Feng10} community suggests a super-stiff symmetry energy. Therefore, much more efforts
need to be done to better determine the nuclear symmetry energy at both supersaturation and
subsaturation densities.

Heavy-ion collisions induced by neutron-rich nuclei as an important tool are commonly used to
study the density dependence of nuclear symmetry energy \cite {Sto86,Cas90,Har96,Ko96,Bao98,
Ditoro99}. Usually, a higher compressive density formed in central heavy-ion collisions with
the softer symmetry energy gets the \rpi ratio to be larger compared to the case of stiffer
symmetry energy. However, the densities formed in heavy-ion collisions always experience a
broad range from subsaturation to supersaturation densities. Therefore, one has to evaluate
the influence of the high-density (low-density) matter phase on observable when probing the
symmetry energy at subsaturation (supersaturation) density using heavy-ion collisions due to
the formation of supersaturation (subsaturation) density matter. Certainly, the influence of
low-density matter phase is inevitable using heavy-ion collisions to probe high-density
symmetry energy due to the densities formed at the final reaction stage is always lower than
the saturation density, therefore, one has to select those of reaction production without
experiencing the final reaction stage such as pre-equilibration neutron-proton ratio.
On the other hand, as shown recently, the pion potential has an opposite effects on \rpi
ratio compared to the effect of symmetry energy on it, and thus decreases the sensitivity of
\rpi ratio to symmetry energy \cite{Hong14,Guo15}. Moreover, the modification of pion production
threshold can even invert the sensitivity of \rpi ratio to symmetry energy \cite{Song15}.
Actually, impact parameter as a factor can also influence the compressive density of participating
region and thus may even invert the sensitivity of \rpi ratio to symmetry energy in peripheral
collisions as mentioned in our recent work about the influence of neutron-skin thickness on the
\rpi ratio in heavy-ion collisions \cite{Wei14}. Therefore, it is necessary to systematically
check the impact parameter dependence of \rpi ratio in probing the symmetry energy using heavy-ion
collisions, and show the corresponding reasons and which energy range of pion does get the
sensitivity of \rpi ratio to symmetry energy reversal. This is the main purpose of the present study.

\begin{figure}[h]
\centerline{\includegraphics[scale=0.34]{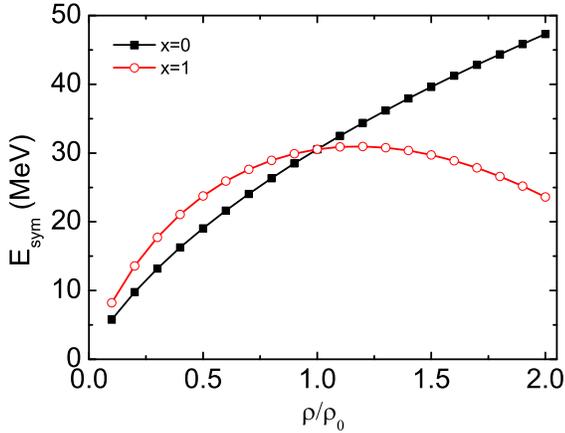}} \caption{The density dependence of nuclear symmetry energy. } \label{esym}
\end{figure}

The present study is based on an isospin-dependent Boltzmann-Uehling-Uhlenbeck (IBUU) transport
model \cite{IBUU}. In this model, an isospin-dependent mean-field is used to model the nuclear
interaction, its expression is defined as follows:
\begin{eqnarray}
U(\rho,\delta ,\vec{p},\tau ) &=&A_{u}(x)\frac{\rho _{-\tau }}{\rho _{0}}%
+A_{l}(x)\frac{\rho _{\tau }}{\rho _{0}}  \notag \\
&+&B(\frac{\rho }{\rho _{0}})^{\sigma }(1-x\delta ^{2})-8\tau x\frac{B}{%
\sigma +1}\frac{\rho ^{\sigma -1}}{\rho _{0}^{\sigma }}\delta \rho
_{-\tau }
\notag \\
&+&\frac{2C_{\tau ,\tau }}{\rho _{0}}\int d^{3}p^{\prime }\frac{f_{\tau }(%
\vec{p}^{\prime })}{1+(\vec{p}-\vec{p}^{\prime })^{2}/\Lambda ^{2}}
\notag \\
&+&\frac{2C_{\tau ,-\tau }}{\rho _{0}}\int d^{3}p^{\prime }\frac{f_{-\tau }(%
\vec{p}^{\prime })}{1+(\vec{p}-\vec{p}^{\prime })^{2}/\Lambda ^{2}}.
\label{MDIU}
\end{eqnarray}%
In the above, $\rho=\rho_n+\rho_p$ is the nucleon number density and $\delta=(\rho_n-\rho_p)/\rho$
is the isospin asymmetry of the nuclear medium; $\rho_{n(p)}$ denotes the neutron (proton) density,
the isospin $\tau$ is $1/2$ for neutrons and $-1/2$ for protons, and $f(\vec{p})$ is the local phase
space distribution function. The expressions and values of the parameters $A_{u}(x)$, $A_{l}(x)$,
$\sigma$, $B$, $C_{\tau ,\tau }$, $C_{\tau ,-\tau }$, and $\Lambda $ can be found in
Refs. \cite{Das03,Che05}, and they lead to the binding energy of $-16$ MeV, incompressibility
$212$ MeV for symmetric nuclear matter, and symmetry energy $E_{sym}(\rho_0)=30.5$ MeV at saturation
density $\rho_0=0.16$ fm$^{-3}$, respectively. While parameter $x$ is used to mimic the different
forms of symmetry energy predicted by various many-body theories without changing any properties
of symmetric nuclear matter and the value of symmetry energy at saturation density \es0. Shown in
Fig. \ref{esym} is the density dependence of symmetry energy with a softer setting $x$=1 and stiffer
one $x$=0.

\begin{figure}[h]
\centerline{\includegraphics[scale=0.34]{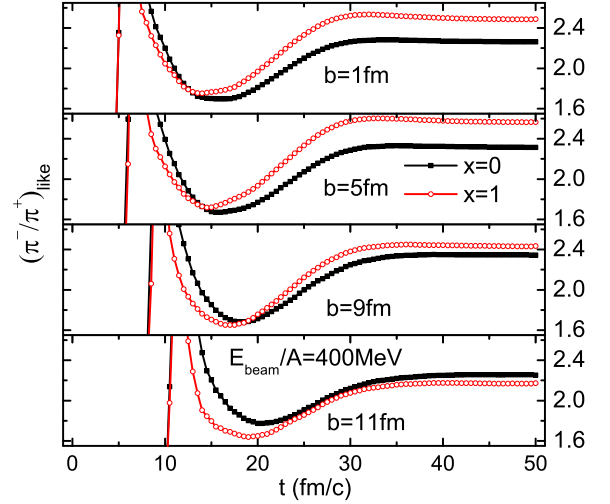}} \caption{The time evolution of (\rpi)$_{like}$ ratio from
central to peripheral Pb+Pb collisions at the beam energy of 400 MeV/nucleon.}\label{pilike}
\end{figure}
\begin{figure}[h]
\centerline{\includegraphics[scale=0.34]{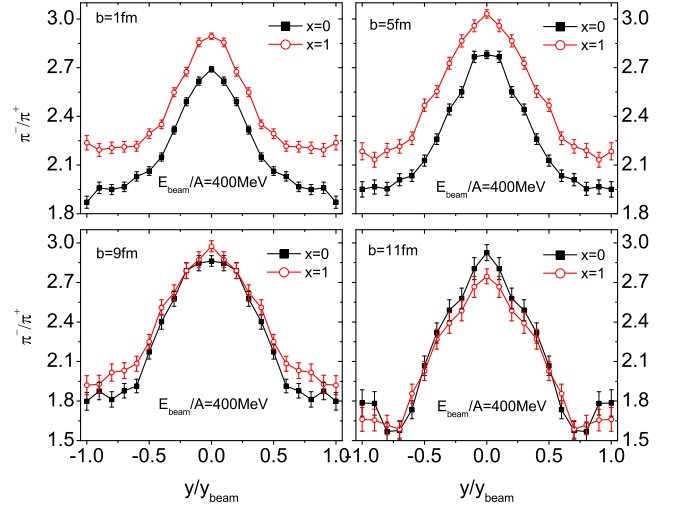}} \caption{The rapidity distribution of \rpi ratio from central
to peripheral Pb+Pb collisions at the beam energy of 400 MeV/nucleon.}\label{rapidity}
\end{figure}
\begin{figure}[h]
\centerline{\includegraphics[scale=0.35]{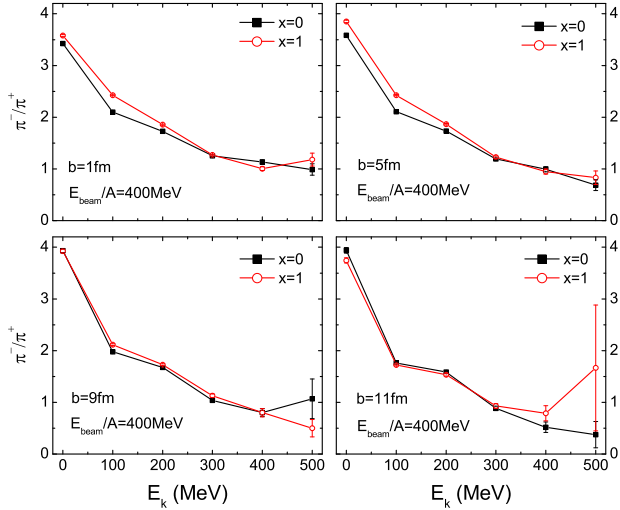}} \caption{The kinematic energy distribution of \rpi
ratio from central to peripheral Pb+Pb collisions at the beam energy of 400 MeV/nucleon.}\label{kinematic}
\end{figure}
\begin{figure}[h]
\centerline{\includegraphics[scale=0.34]{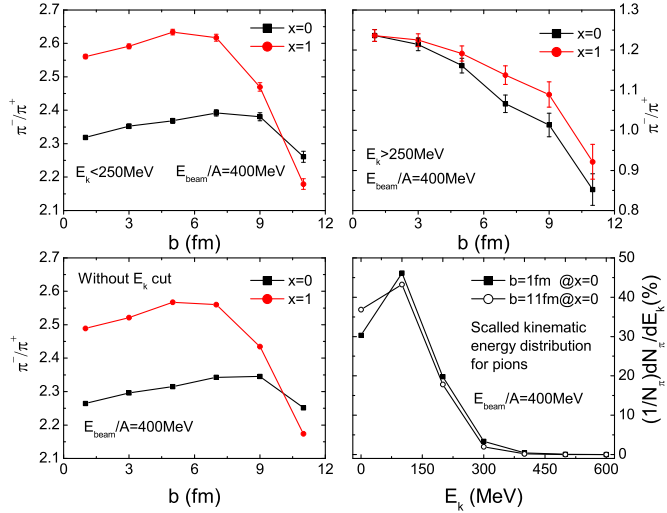}} \caption{The impact parameter dependence of
\rpi ratio formed by pions with kinematic energy lower than 250MeV, larger than 250MeV, and without
any kinematic energy cut from Pb+Pb collisions at the beam energy of 400 MeV/nucleon; and the
kinematic energy distribution of pion number percentage from Pb+Pb collisions with symmetry energy parameter
$x$=0 at two impact parameters of 1fm and 11fm and the beam energy of 400MeV/nucleon.}\label{group}
\end{figure}
\begin{figure}[h]
\centerline{\includegraphics[scale=0.34]{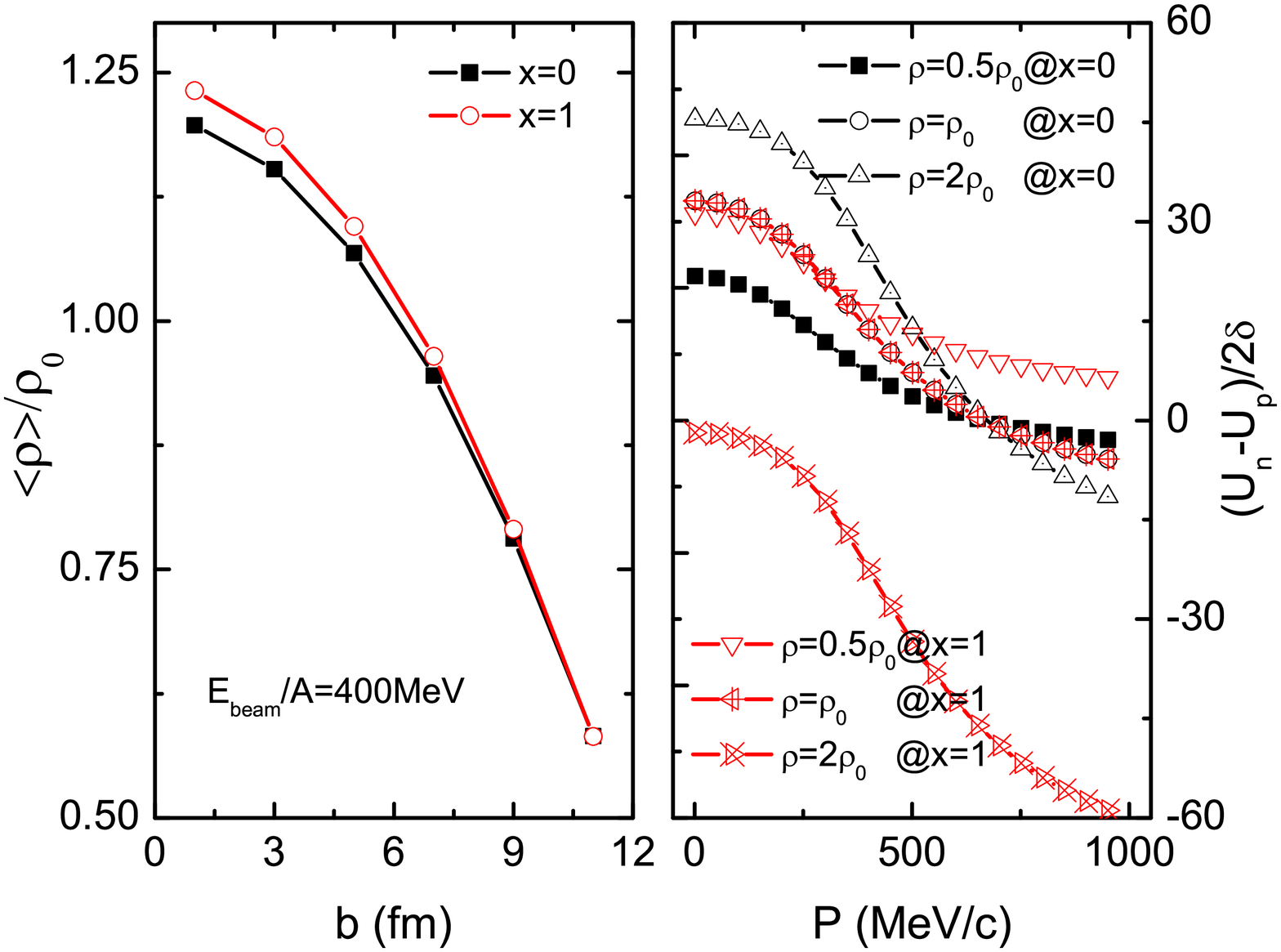}} \caption{The average density of participating
region over the whole reaction time from central to peripheral Pb+Pb collisions at the beam
energy of 400 MeV/nucleon; and the momentum dependence of symmetry potential at different nuclear
matter density.}\label{density}
\end{figure}
Now let's check the impact parameter dependence of \rpi ratio in probing the symmetry energy. Within
the IBUU transport model for heavy-ion collision at the intermediate energy, almost all the pions are
produced from the decay of $\Delta$(1232) resonances. Therefore, the dynamic pion ratio, i.e.,
(\rpi)$_{\rm like}$, can be defined as
\begin{equation}\label{ratio}
(\pi^{-}/\pi^{+})_{\rm like}\equiv
\frac{\pi^{-}+\Delta^{-}+\frac{1}{3}\Delta^{0}}
{\pi^{+}+\Delta^{++}+\frac{1}{3}\Delta^{+}}.
\end{equation}
Due to all the $\Delta$ resonances will eventually decay at the final reaction stage, it is thus the (\rpi)$_{\rm like}$
ratio will naturally become the \rpi ratio. Shown in Figs. \ref{pilike} and \ref{rapidity} are the time evolution
of (\rpi)$_{\rm like}$ ratio and rapidity distribution of \rpi ratio from central to peripheral Pb+Pb collisions
at the beam energy of 400 MeV/nucleon. Similar to previous results \cite{Xiao09,Feng10}, the dynamic
(\rpi)$_{\rm like}$ ratio and final \rpi ratio are more sensitive to the symmetry energy at the central
heavy-ion collision compared to the case of peripheral heavy-ion collision, and larger with a softer symmetry energy
setting $x$=1 compared to the case of the stiffer setting $x$=0. However, it can be seen that no matter the dynamic
(\rpi)$_{\rm like}$ ratio or the rapidity distribution of final \rpi ratio on the symmetry energy show a transition
from central to peripheral collisions, i.e., the stiffer symmetry energy leads to a larger \rpi ratio in peripheral
collisions while the softer symmetry energy obviously gets this ratio to be larger in central collisions. On the
other hand, it is well known that the increasing of impact parameter will directly change the participant numbers,
and thus the pion multiplicities and kinematic energy distribution. Therefore, a natural question is which energy
range of pion does invert the sensitivity of \rpi ratio to symmetry energy from central to peripheral collisions.
To this end, we show in Fig. \ref{kinematic} the kinematic energy distribution of \rpi ratio with different impact parameter.
In general, it is similar to above observation the sensitivity of \rpi ratio to symmetry energy is decreasing at lower
kinematic energy as increasing the impact parameter especially from midcentral to peripheral collisions, and even shows
a opposite sensitivity in very peripheral collisions. However, for the pion ratio at larger kinematic energy, its value
with the softer symmetry energy is also larger even in very peripheral collisions albeit with a larger error bar.
This naturally gets us to look at the impact parameter dependence of \rpi ratio formed by less energetic pions
and more energetic pions, separately. For this purpose, we take empirically a kinematic energy cut of 250MeV and
classify pions into less energetic and more energetic groups. Shown in upper panel of Fig. \ref{group} are the
impact parameter dependence of \rpi ratio formed by less energetic pions and more energetic pions, respectively.
It is seen that the transition of sensitivity of \rpi ratio to symmetry energy is mainly from less energetic
pions, i.e., the softer symmetry energy gets the less energetic pions to form a smaller \rpi ratio in peripheral
collisions while these pions generate a larger \rpi ratio in central collisions. Certainly, the softer symmetry energy
also leads more energetic pions to form a larger \rpi ratio in peripheral collisions. Nevertheless, due to most of
pions are less energetic at this beam energy as shown in the right plot at lower panel of Fig. \ref{group}, thus the
behaviour of \rpi ratio formed by all pions without any kinematic energy cut is almost similar to those formed by
less energetic pions as shown in the left plot at lower panel of Fig. \ref{group}.

Now let us to show the reason of pion ratio transition in probing the symmetry energy from central to peripheral
collisions. To this end, we show the average density of participating region over the whole reaction time in the
left panel of Fig. \ref{density} from central to peripheral collisions. It can be found that almost all the pions
are produced at supersaturation density at central heavy-ion collisions but subsaturation density at peripheral
collisions. On the other hand, from the symmetry potential in right panel of Fig. \ref{density} and symmetry
energy in Fig. \ref{esym}, it can be found that the stiffer symmetry energy with parameter $x$=0 generates a
larger symmetry energy and a larger symmetry potential when the density of participating region is higher than
the normal density, thereby generating a stronger repulsive effects for neutrons but attractive effects for
protons and thus leading to a smaller \rpi ratio in central collisions. On the contrary, when the density of
participating region is lower than the normal density, the stiffer symmetry energy with parameter $x$=0
corresponds to a smaller symmetry energy and a smaller symmetry potential compared to the case of the softer
symmetry energy with parameter $x$=1, and then naturally generating a larger \rpi ratio in peripheral collisions.
This implies the \rpi ratio as a probe of high-density symmetry energy effective only in central at most
to midcentral collisions, thereby avoiding the possible information of low-density symmetry energy
carried in \rpi ratio from peripheral collisions.

In summary, we have carried out an investigation about the impact parameter dependence of \rpi
ratio in probing the nuclear symmetry energy using heavy-ion collision. Within an isospin-dependent
transport model, the Pb+Pb collisions are performed with different impact parameter at a beam energy
of 400 MeV/nucleon. It is shown that the sensitivity of \rpi ratio on symmetry energy has a transition
from central to peripheral collisions due to the less energetic pions measure the high-density symmetry
energy in central collisions but the low-density symmetry energy in peripheral collisions. Therefore,
we suggest the \rpi signature as a high-density symmetry probe effective only in central at most to
midcentral collisions. Certainly, other effects such as pion production threshold and energy conservation
and pion potential, which are not considered in the present study, can also influence significantly the
sensitivity of \rpi ratio to symmetry energy as shown in others \cite{Hong14,Guo15,Song15}. \\

\noindent{\textbf{Acknowledgements}} \\
This work is supported by the National Natural Science Foundation of China under grant No.11405128
and Xi'an Science and Technology Planning Project No. CXY1352WL29.

\end{CJK*}

\end{document}